\documentclass[a4paper, superscriptaddress, twocolumn, prl, showpacs, longbibliography]{revtex4-1}

\usepackage{dcolumn}
\usepackage{bm}
\usepackage[colorlinks=true,citecolor=blue,linkcolor=blue,pdfhighlight =/O]{hyperref}
\usepackage{graphicx}
\usepackage{amsmath}
\usepackage{dsfont}
\usepackage{mathtools}
\usepackage[dvipsnames]{xcolor}
\usepackage{soul}
\usepackage[normalem]{ulem}

\newcommand{\be}{\begin{equation}}
\newcommand{\ee}{\end{equation}}

\hypersetup{urlcolor=blue}

\begin{document}
\title{Quantum confinement suppressing electronic heat flow below\\ the Wiedemann-Franz law}
\author{D. Majidi}
\affiliation{\mbox{Univ.} Grenoble Alpes, CNRS, Grenoble INP, Institut N\' eel, 25 rue des Martyrs, Grenoble, France}

\author{M. Josefsson}
\affiliation{NanoLund and Solid State Physics, Lund University, Box 118, 22100 Lund, Sweden}

\author{M. Kumar}
\affiliation{NanoLund and Solid State Physics, Lund University, Box 118, 22100 Lund, Sweden}

\author{M. Leijnse}
\affiliation{NanoLund and Solid State Physics, Lund University, Box 118, 22100 Lund, Sweden}

\author{L.~Samuelson}
\affiliation{NanoLund and Solid State Physics, Lund University, Box 118, 22100 Lund, Sweden}

\author{H. Courtois}
\affiliation{\mbox{Univ.} Grenoble Alpes, CNRS, Grenoble INP, Institut N\' eel, 25 rue des Martyrs, Grenoble, France}

\author{C. B. Winkelmann}
\affiliation{\mbox{Univ.} Grenoble Alpes, CNRS, Grenoble INP, Institut N\' eel, 25 rue des Martyrs, Grenoble, France}

\author{V. F. Maisi}
\affiliation{NanoLund and Solid State Physics, Lund University, Box 118, 22100 Lund, Sweden}
\date{\today}
\date{\today}

\begin{abstract}
 The Wiedemann-Franz law states that the charge conductance and the electronic contribution to the heat conductance are proportional. This sets stringent constraints on efficiency bounds for thermoelectric applications, which seek for large charge conduction in response to a small heat flow. We present experiments based on a quantum dot formed inside a semiconducting InAs nanowire transistor, in which the heat conduction can be tuned significantly below the Wiedemann-Franz prediction. Comparison with scattering theory shows that this is caused by quantum confinement and the resulting energy-selective transport properties of the quantum dot. Our results open up perspectives for tailoring independently the heat and electrical conduction properties in semiconductor nanostructures.
\end{abstract}

\maketitle

In conductors, a higher electrical conductance $G$ is generally associated to a correspondingly higher heat conductance $\kappa$. The Wiedemann-Franz (WF) law indeed stipulates that at a given temperature $T$, the ratio defined as $L=\kappa /GT$ is constant and equal to the Lorenz number $L_0=(\pi^2/3)(k_B/e)^2$. The connection of the two quantities arises from the fact that both charge and heat are carried by the same particles, and has been experimentally verified to hold down to the scale of single-atom and molecule contacts \cite{cui2017,mosso2017}. Deviations indicate departures from Fermi liquid physics \cite{benenti2017} such as found in superconductors \cite{ambegaokar1964}, correlated electron systems \cite{lee2017}, Majorana modes \cite{pan2021} or viscous electron flow \cite{crossno2016}. 
In quantum nanodevices, Coulomb interaction and charge quantization in metallic nanoislands were also shown to lead to departures from the WF law \cite{kubala2008,dutta2017,sivre2018}.

\begin{figure}[t]
	\includegraphics[width=1\columnwidth]{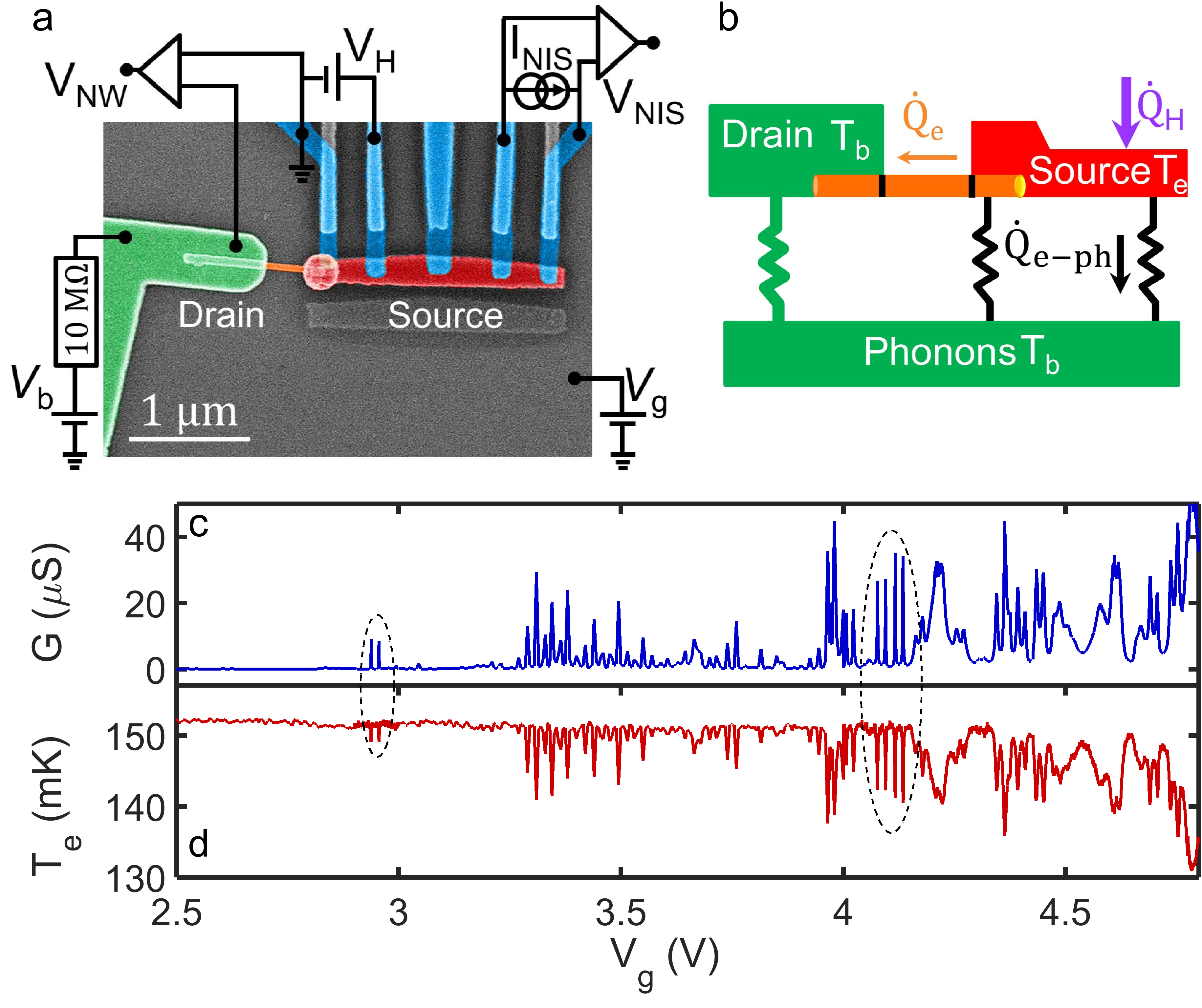}
	\caption{Heat transport experiment through an InAs nanowire device. (a) False-colored scanning electron micrograph of the device. The drain electrode, the source island and the nanowire are colored in green, red and orange, respectively. Five superconducting aluminum leads (light blue) are connected to the source island for heating the source side and measuring its electronic temperature. Thermometry is performed by measuring the voltage $V_\mathrm{NIS}$ at a fixed floating current bias $I_\mathrm{NIS}$. (b) Heat balance diagram, which includes the applied power to the source island, $\dot{Q}_H$; the heat escaping due to electron-phonon coupling, $\dot{Q}_{\rm e-ph}$; and the electronic heat flow along the nanowire, $\dot{Q}_e$. (c) Electrical conductance at thermal equilibrium and (d) temperature response $T_e$ of the source island with heating power of $\dot{Q}_H= 16$ fW as a function of the back gate voltage $V_{g}$. The dashed ellipses highlight resonances that will be studied in more detail. All measurement are taken at a bath temperature $T_b=100$ mK.}
	\label{fig:figure1}
\end{figure}

In semiconducting materials the WF law is notoriously well obeyed for the electronic contribution to heat conductance, including semiconducting nanostructures displaying transport in the quantum Hall state \cite{jezouin2013,banerjee2018}.
This property imposes severe limitations for instance in thermoelectrics, for which it is desirable to maximize the charge flow while minimizing that of heat. The most common figure of merit for thermoelectric conversion, $ZT$, is indeed directly proportional to $L^{-1}$. Nevertheless, semiconducting nanostructures can display adjustable and strongly energy-selective transport processes, which could also lead to breaking the WF law, even in the absence of interaction effects. This can be provided for instance by the quantization of the energy levels in a single-quantum-dot junction, allowing for an adjustable narrow transmission window in energy. Although theory has predicted a vanishing $L/L_0$ for weakly tunnel-coupled quantum dots at low temperature \cite{zianni2007, krawiec2006thermoelectric, krawiec2007thermoelectric, murphy2008, tsaousidou2010thermoelectric, erdman2017thermoelectric}, it was experimentally shown that higher-order effects restore a significant electronic heat leakage \cite{dutta2020}. The validity of the WF law in a single-quantum-dot device has however not yet been quantitatively investigated because of the difficulty in measuring the extremely small heat currents.

In this work we investigate heat flow in a quantum dot formed in an InAs nanowire grown by chemical beam epitaxy \cite{bjork2004,fasth2007direct}. Such nanowires have been widely studied for their promising thermoelectric properties \cite{wu2013, roddaro2013, chen2018, svilans2018, prete2019}. It was recognized that the formation of quantum-dot-like states in nanowires can lead to a large enhancement of the thermopower, well beyond expectations from 1D models \cite{wu2013}. Such quantum dots can be produced either by inserting controlled InP tunnel barriers, or simply by the inherent electrostatical nonuniformities at low carrier density. They recently allowed experimentally testing the Curzon-Ahlborn limit of thermoelectric conversion efficiency at maximum power \cite{josefsson2018}. Although entering directly in the thermoelectric efficiencies, the electronic heat conductance of such devices is in general not measured independently. Because at temperatures above a few degrees Kelvin, the thermal transport properties of InAs nanowires are known to be strongly dominated by phonons \cite{matthews2012}, the {\it electronic} heat conductance of InAs can only be experimentally probed at milliKelvin temperatures.

The experimental device is an InAs nanowire of 70 nm diameter, back-gated from the degenerately doped silicon substrate at a potential $V_g$ and electrically connected on one side to a large gold contact named {\it drain} from hereon (\mbox{Fig.  \ref{fig:figure1}}a). The nanowire conductance $dI/dV_\mathrm{NW}$ is measured using a voltage division scheme as pictured in \mbox{Fig.  \ref{fig:figure1}a}, involving a 10 M$\Omega$ bias resistor. The other side (the {\it source}) consists in a few-micrometer-long normal metallic island, connected by five superconducting aluminum leads. The leftmost of these in \mbox{Fig. \ref{fig:figure1}}a is in direct ohmic contact with the source island. This allows measuring directly the nanowire linear charge conductance $G(V_g)$, as shown in \mbox{Fig. \ref{fig:figure1}}c. In agreement with previous reports on similar structures \cite{wu2013}, the nanowire conduction is pinched off below $V_g\sim 3$ V. Near pinch off, the conductance displays sharp resonances, which indicate that the nanowire conduction bottleneck at vanishing charge carrier densities will be provided by a quantum dot forming in the part of the nanowire that is not below the metallic contacts (Fig. \ref{fig:figure1}c). Although "unintentional" (in contrast with epitaxially engineered quantum dots \cite{bjork2004,dick2010crystal}), these quantum dots display well-defined level quantization $\delta \varepsilon$, tunnel coupling strengths $\gamma_{s,d}$ and charging energies $E_c$ all three significantly larger than $k_BT$. Here, $k_B$ is the Boltzmann constant and $T$ the experimental working temperature, which is set to $T_b=100$ mK at equilibrium. Details of the charge conductance properties, which we extract from full $dI/dV_\mathrm{NW}(V_\mathrm{NW},V_g)$
differential conductance maps, are found in the  Supp. Info. file.

The other four aluminum leads to the source are in contact via tunnel barriers. Such superconductor-insulator-normal metal (NIS) junctions are well-known to provide excellent electron heaters and thermometers in low temperature experiments \cite{giazotto2006opportunities}. Because at mK temperatures both the electron-phonon coupling in metals and the heat conductance of superconductors are very low, the source island electrons are thermally well insulated, such that the heat flow through the nanowire significantly contributes to the source island's heat balance. This is seen in \mbox{Fig. \ref{fig:figure1}d}, in which a constant heating power ${\dot Q}_H= 16$ fW is provided to the source island via a voltage $V_H$ applied on one tunnel lead. As the gate potential is swept, the variations of the source island electron temperature $T_e$ are strikingly anticorrelated to variations of $G$. The heat balance of our device is schematised in Fig. \ref{fig:figure1}b. Because the source island is overheated with respect to its environment, the gradual opening of electronic conduction channels in the InAs nanowire leads to increased heat flow out of the source island, and thus a lowering of $T_e$.

In the remainder of this work, we investigate quantitatively the nanowire heat conductance properties, and compare it to the predictions of both the WF law and the Landauer-B\"uttiker scattering theory \cite{sivan1986multichannel}. To this end, it is very insightful to go beyond linear response in $\Delta T=T_e-T_b$, and we thus measure at every gate voltage the full relation ${\dot Q}_H(T_e,V_g)$ between the Joule power ${\dot Q}_H$ applied to the source and its internal equilibrium electronic temperature $T_e$. Details of the determination of ${\dot Q}_H$ are described in the Supp. Info. file.

An important issue in the determination of electronic heat flow is the proper identification of the parasitic heat escape via other channels, such as electron-phonon coupling \cite{giazotto2006opportunities}. Unless the latter can be neglected \cite{jezouin2013}, the comparison to a reference, at which the electronic heat conductance is either assumed to be known \cite{dutta2017}, or negligible, is required. We define ${\dot Q}_H(T_e,0)$ measured deep in the insulating regime as an experimental reference which contains all heat escape channels out of the source island other than mediated by the nanowire charge carriers. We stress that this choice does not rely on any thermal model and we furthermore consider the gate-dependent part of the heat balance, defined as ${\dot Q}(T_e,V_g)={\dot Q}_H(T_e,V_g)-{\dot Q}_H(T_e,0)$. The magnitude and temperature dependence of ${\dot Q}_H(T_e,0)$ is in good agreement with estimates for the electron-phonon coupling in the metallic parts of the source (see Supp. Info.). Surprisingly, we observe that ${\dot Q}(T_e,V_g)$ is slightly gate dependent even before the conducting state sets on. This is readily visible as a slightly negative slope of the $T_e(V_g)$ baseline in Fig. \ref{fig:figure1}d. We thus conclude on a minute yet measurable and smoothly gate-dependent contribution to the source electron-phonon coupling from the part of the nanowire below the source, which calls for defining in addition a local reference, as discussed below.

\begin{figure}[t]
	\includegraphics[width=1\columnwidth]{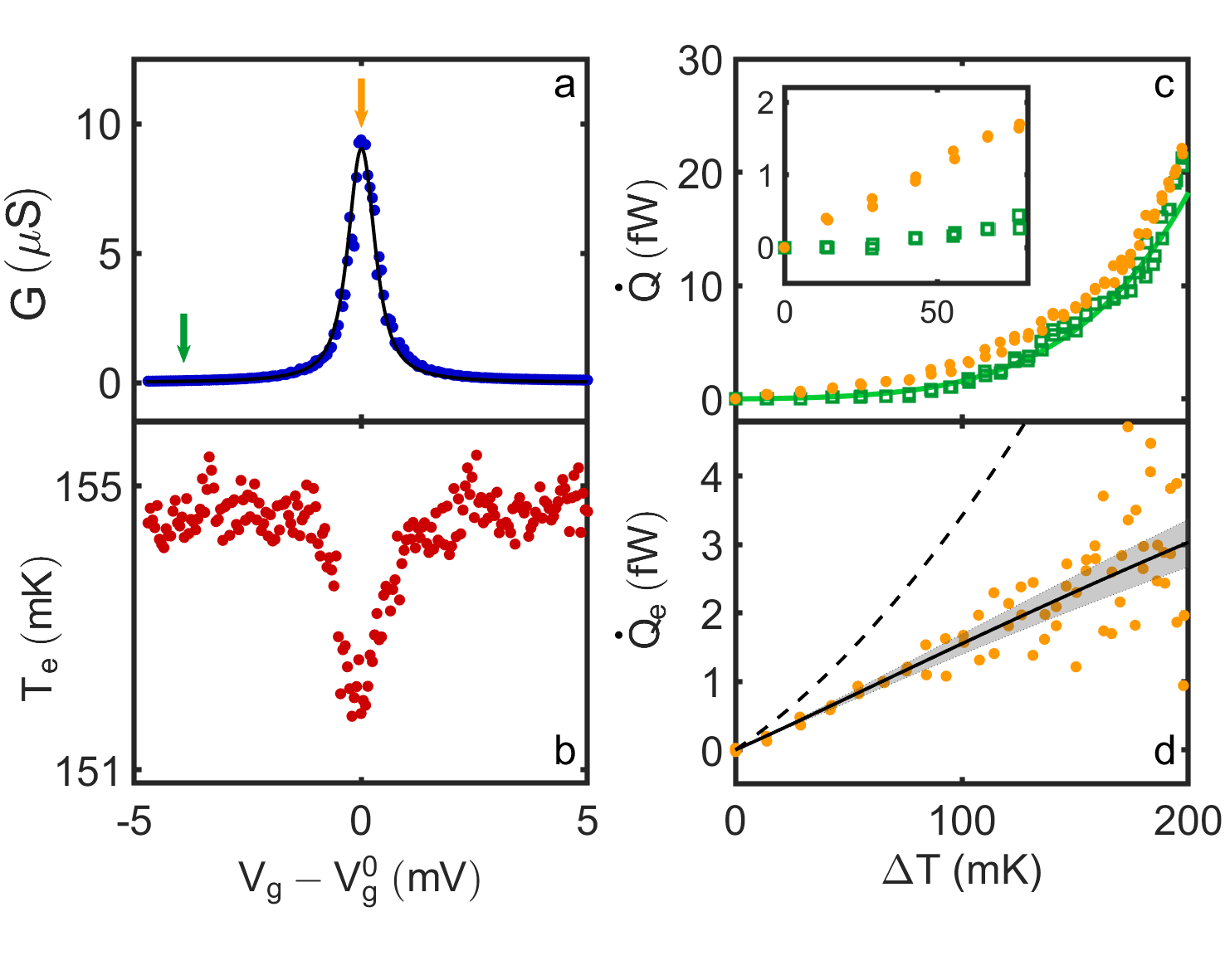}
	\vspace{-0.9cm}
	\caption{Heat transport near an isolated conductance resonance. (a) Linear charge conductance around $V_g^0= 2.938$ V. The black line is a fit using scattering theory. (b) Source temperature $T_e$ as a function of $V_{g}$, with a constant applied power $\dot{Q}_H= 16$ fW. (c) Full heat balance curve ${\dot Q}(T_e,V_g)$ on (green squares) and off (orange bullets) the transport resonance, as indicated by the arrows in (a). The green line presents a fit using $\dot Q = \beta (T_e^6-T_b^6)$ with $\beta = 35 \pm 5$ pW/K$^6$. The inset highlights the electronic contribution, dominating at small temperature difference at the resonance. (d) Difference of the two data sets in c, displaying the purely electronic heat transport contribution $\dot{Q}_e$. The dashed and the full lines are the predictions from the WF law and scattering transport theory, respectively. The grey shaded area indicates the uncertainty of the scattering theory calculation, due to the determination of the gate coupling lever arm.}
	\label{fig:figure2}
\end{figure}

The very first conduction resonance, visible in Fig. \ref{fig:figure1}c,d and Fig. \ref{fig:figure2}a,b at $V_g^0=2.938$ V, is ideally suited for a {\it local} differentiation of the electronic heat conductance $\dot{Q}_e$ through the nanowire over the smooth e-ph background contribution $\dot{Q}_\mathrm{e-ph}$ of the source side. At gate voltages $|\Delta V_g |\geq 3$ mV away from the conduction resonance at $V_g^0$, the heat flow ${\dot Q}(T_e,V_g)$ is constant, within noise, although the charge conductance $G$ still varies. After differentiation of the heat balance on and off resonance (Fig. \ref{fig:figure2}c), one is thus left with the quantity of interest, the {\it electronic} heat flow through the nanowire at resonance, ${\dot Q_e(T_e,V_g^0)}={\dot Q}(T_e,V_g^0)-{\dot Q}(T_e,V_g^0+\Delta V_g)$. We stress that this additional background subtraction does not rely on any modeling of the heat balance, such as electron-phonon coupling. As seen in Fig. \ref{fig:figure2}d and already visible in the inset of Fig. \ref{fig:figure2}c, ${\dot Q}_e$ at $V_g^0$ displays a strikingly linear dependence on $\Delta T$. We see that the heat conductance $\kappa_e=\partial {\dot Q}_e/\partial T$, that is the initial slope in Fig. \ref{fig:figure2}d, differs quantitatively from the WF prediction by a factor $L/L_0\approx0.65\pm0.1$. Further, beyond linear response, the temperature dependence qualitatively deviates from the parabolic law expected from WF (dashed line).

For a theoretical description beyond the WF law, we use a Landauer-B\"uttiker non-interacting model, with an energy-dependent transmission ${\mathcal T}(E)$. We write the associated charge and heat currents, respectively as
\begin{equation}
    I=\frac{2e}{h}\int_{-\infty} ^\infty {\mathcal T}(E)\  \Delta f\  dE
    \label{Eq:I}
\end{equation}
and
\begin{equation}
    {\dot Q}_e=\frac{2}{h}\int_{-\infty} ^\infty (E-\mu_s) \  {\mathcal T}(E)\ \Delta f \  dE,
    \label{Eq:Q}
\end{equation}
with $\Delta f$ the difference in the source and drain energy distributions, and $\mu_s$ the source island chemical potential \cite{sivan1986multichannel,davies1998physics}. 
The linear charge and heat conductances are then obtained as $G=\partial I/\partial V_\mathrm{NW}$ and $\kappa_e=\partial {\dot Q}_e/\partial (\Delta T)$, respectively, with $\Delta T=T_e-T_b$. 
We model each resonance as a discrete energy level coupled to the source and drain reservoirs. 
We then deduce the transmission function ${\mathcal T}(E)$ by fitting the calculated gate-dependent charge conductance $G(V_g)$ to the data. The accurate determination of ${\mathcal T}(E)$ requires accurately estimating independently the tunnel couplings and the gate lever arm, as both affect similarly the resonance widths. This is described in detail in the Supp. Info. file. On a technical note, we stress that the above theoretical expression of $\kappa_e$ assumes open-circuit conditions, that is, no net particle current. For all heat conductance experiments the nanowire was biased in series with a 10 M$\Omega$ resistor at room temperature. Because we only consider data at gate voltages at which $G$ is significantly larger than $(10\ \mathrm{M\Omega})^{-1} = 0.1\,\mu$S, applying $V_b=0$ is then equivalent to imposing open circuit conditions.

With the above analysis, the Landauer-B\"uttiker theoretical ${\dot Q_{\rm e}(T_e,V_g)}$ follows directly. As seen in \mbox{Fig. \ref{fig:figure2}d} (solid black line), the agreement with the experimental data is very good, with no adjustable parameters, reproducing the observed approximately linear dependence on $\Delta T$. The grey shaded region accounts for the uncertainties in the determination of ${\mathcal T}(E)$. The violation of the WF law observed here is therefore accurately described by a non-interacting scattering transport picture.

Intuitively, the deviation from WF at resonance can be understood as stemming from the energy selectivity of the device transmission, so that tunneling electrons carry an energy bound by $\sim \gamma_s+\gamma_d$, thus suppressing heat exchange at zero net charge current.
Together with a large Seebeck coefficient \cite{wu2013,roddaro2013}, this reduction of heat conductance without suppressing particle conductance makes the quantum dot junction potentially the "best thermoelectric" as theorised by Mahan and Sofo \cite{mahan1996}. With increasing tunnel couplings, the transmission function ${\mathcal T}(E)$ is broadened and the energy selectivity is gradually lost.

\begin{figure}[t]
\centering
	\includegraphics[width=0.75\columnwidth]{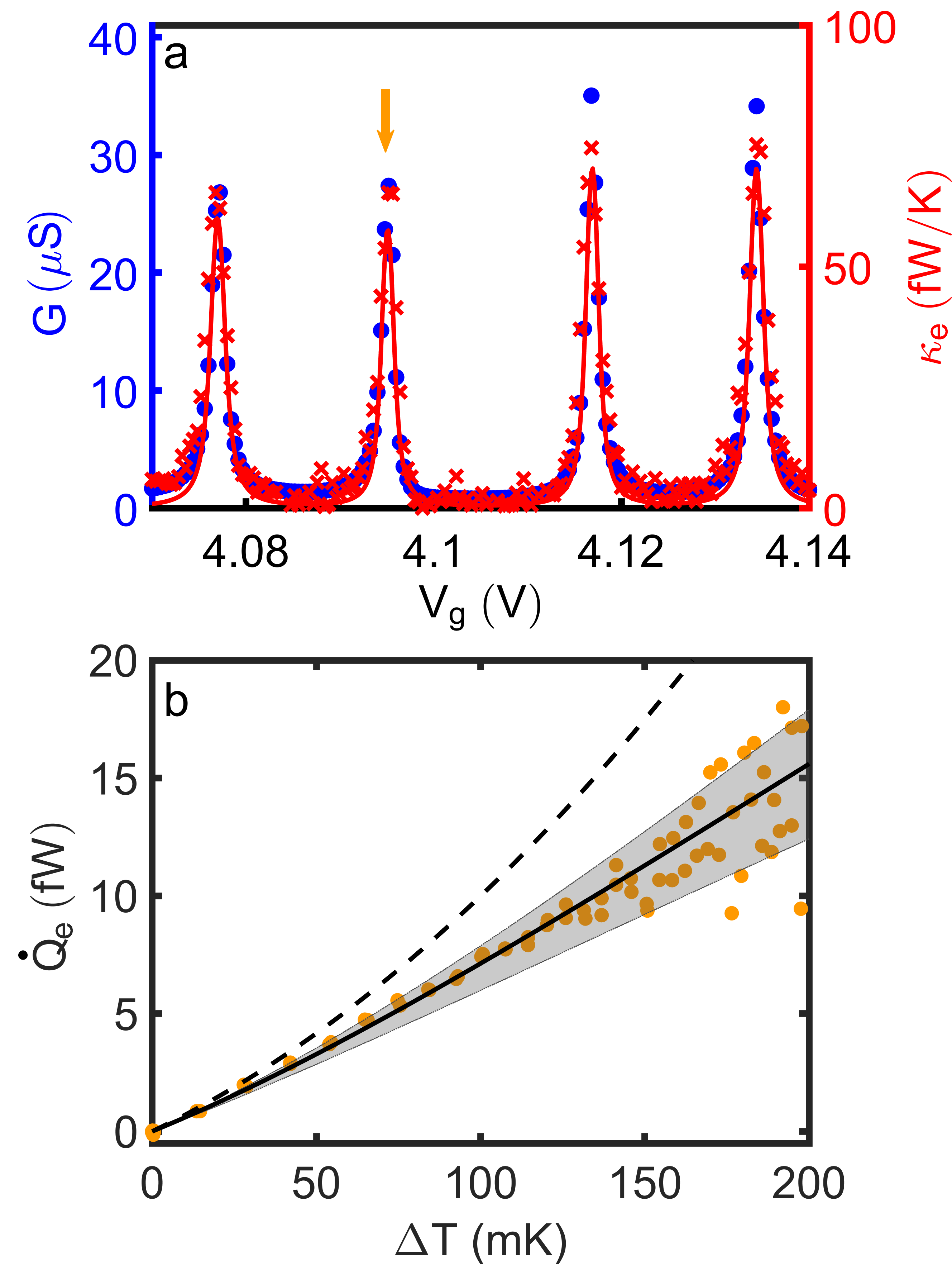}
	\caption{Heat versus charge transport at higher transmissions. (a) Heat (red crosses, right vertical scale) and charge (blue bullets, left vertical scale) conductance resonances at higher transmissions. The ratio of both vertical scales is set to $T_bL_0$, such that superimposed curves are indicative of the WF law being valid. The red line is the calculated $\kappa_e$ from scattering transport theory. The $L/L_0$ for the four peaks are 0.99, 0.97, 0.87 and 0.90 $(\pm 0.05)$ from left to right. (b) ${\dot Q_{\rm e}(T_e)}$ curve taken at the conduction resonance at $V_g=4.095$ V (arrow in (a)). The dashed and the full lines are the predictions from the WF law and scattering transport theory, respectively. The grey shaded area indicates the uncertainty of the scattering theory calculation, due to the determination of the gate coupling lever arm.}
	\label{fig:figure3}
\end{figure} 

We exemplify this gradual recovery of the WF law by studying the heat flow close to the conductance resonances observed at a larger gate voltage $V_g$.
While at $V_g\approx 2.9$ V, a ratio $(\gamma_s+\gamma_d)/k_BT_b\approx 7$ placed the device in the intermediate coupling regime, still displaying sizable energy selectivity (Fig. \ref{fig:figure2}), at $V_g\approx 4.1$ V the tunnel couplings are about a factor 2.5 larger (Fig. \ref{fig:figure3}a). We therefore expect a gradual transition to a WF-like heat conductance. This is seen in Fig. \ref{fig:figure3}a, where we superimpose the experimentally determined $G$ and $\kappa_e$ on a vertical scale connecting both quantities via the WF law, that is, $\kappa_e=GT_bL_0$. At the charge degeneracy points (conduction resonances) we observe that the dimensionless reduced heat conductance $L/L_0$ is now very close to, or barely below 1. Moving away from the conductance peak, $G$ and $\kappa_e$ also superimpose nearly exactly, within noise, as also expected from a scattering transport calculation with a now broader ${\mathcal T}(E)$ (line). Observing a sizable deviation from WF requires going beyond linear response (Fig. \ref{fig:figure3}b) \cite{lopez2013}, where the experimental data and the scattering transport calculation remain nevertheless now much closer to the WF law. The main conclusion we draw here is that for increasing tunnel couplings, the scattering theory still describes the experimental data very accurately and over a large temperature difference range. In the linear response regime (small $\Delta T$), the WF law and scattering theory yield convergent predictions.

Moving to yet larger gate voltages ($V_g>4.5$ V) and thus electronic transmissions, the charge conductance no longer vanishes in between conduction resonances, impeding the identification of a clear cut local reference ${\dot Q_{\rm e-ph}(T_e)}$. This prevents a quantitative separation of the electronic heat flow through the nanowire from the e-ph contribution.

At the lower gate voltages, we however can estimate the e-ph coupling induced by adding carriers to the nanowire segment below the source. This is precisely captured by the off-resonance $\dot{Q}(T_e)$ shown by the green line in Fig.~\ref{fig:figure2}c, which follows a power law $ \propto (T_e^6-T_b^6)$. Interestingly, this leads to an e-ph coupling constant comparable to that of a metal, in spite of the electron density being several orders of magnitude smaller. This finding is consistent with the strong e-ph coupling found in InAs above 1 Kelvin \cite{matthews2012} possibly due to piezoelectricity \cite{prasad2004} and/or a lateral-confinement-enhanced peaked density of states \cite{sugaya2002}. We observe the e-ph contribution to change linearly with $V_g$ (see associated plot and analysis in the Supp. Info. file) implying that the e-ph coupling constant is proportional to the charge carrier density.

In summary, our study reveals large conjunct evolution in the thermal and charge conductances of an InAs nanowire near pinch off. Around conductance resonances in the quantum dot regime of the nanowire, the heat conductance is significantly lower than expected from the WF law, with $\kappa_e/(GTL_0)$ reaching $0.65$ in the intermediate coupling regime, in good agreement with a scattering transport calculation. As anticipated by theory \cite{mahan1996}, this establishes experimentally the huge potential of semiconductor nanowires and more generally quantum dot transistors, as promising high-figure-of-merit thermoelectrics. 
A fascinating open question resides in the role played by electron interactions \cite{dutta2017} that may lead to deviations from the here-employed scattering transport picture away from resonances.

\begin{acknowledgments}
We acknowledge insightful discussions with Denis Basko, Heiner Linke, Nicola Lo Gullo, Jukka Pekola, and Peter Samuelsson. This work received support from the European Union under the Marie Sklodowska-Curie Grant Agreement No. 766025, the Swedish Research Council and NanoLund. V.M. made the devices and performed the experiments. D.M. analysed the data, with help from all authors. M.K. and L.S. grew the nanowires. M.J. and M.L. performed the theoretical calculations. All authors contributed to the interpretation of the data and writing the manuscript.   All data described here will be made publicly available on Zenodo.
\end{acknowledgments}

\bibliography{mybib}

\begin{thebibliography}{10}
\expandafter\ifx\csname url\endcsname\relax
  \def\url#1{\texttt{#1}}\fi
\expandafter\ifx\csname urlprefix\endcsname\relax\def\urlprefix{URL }\fi
\expandafter\ifx\csname href\endcsname\relax
  \def\href#1#2{#2} \def\path#1{#1}\fi

\bibitem{cui2017}
L.~Cui, W.~Jeong, S.~Hur, M.~Matt, J.~C. Kl{\"o}ckner, F.~Pauly, P.~Nielaba,
  J.~C. Cuevas, E.~Meyhofer, P.~Reddy, Quantized thermal transport in
  single-atom junctions, Science 355~(6330) (2017) 1192--1195.

\bibitem{mosso2017}
N.~Mosso, U.~Drechsler, F.~Menges, P.~Nirmalraj, S.~Karg, H.~Riel, B.~Gotsmann,
  Heat transport through atomic contacts, Nat. Nanotechnol. 12~(5) (2017)
  430--433.

\bibitem{benenti2017}
G.~Benenti, G.~Casati, K.~Saito, R.~S. Whitney, Fundamental aspects of
  steady-state conversion of heat to work at the nanoscale, Phys. Rep. 694
  (2017) 1--124.

\bibitem{ambegaokar1964}
V.~Ambegaokar, L.~Tewordt, Theory of the electronic thermal conductivity of
  superconductors with strong electron-phonon coupling, Phys. Rev. 134~(4A)
  (1964) A805.

\bibitem{lee2017}
S.~Lee, et~al., Anomalously low electronic thermal conductivity in metallic
  vanadium dioxide, Science 355~(6323) (2017) 371--374.

\bibitem{pan2021}
H.~Pan, J.~D. Sau, S.~Das~Sarma, Three-terminal nonlocal conductance in
  {M}ajorana nanowires: Distinguishing topological and trivial in realistic
  systems with disorder and inhomogeneous potential, Phys. Rev. B 103~(1)
  (2021) 014513.

\bibitem{crossno2016}
J.~Crossno, et~al., Observation of the {D}irac fluid and the breakdown of the
  {W}iedemann-{F}ranz law in graphene, Science 351~(6277) (2016) 1058--1061.

\bibitem{kubala2008}
B.~Kubala, J.~K{\"o}nig, J.~P. Pekola, Violation of the {W}iedemann-{F}ranz law
  in a single-electron transistor, Phys. Rev. Lett. 100~(6) (2008) 066801.

\bibitem{dutta2017}
B.~Dutta, J.~T. Peltonen, D.~S. Antonenko, M.~Meschke, M.~A. Skvortsov,
  B.~Kubala, J.~K{\"o}nig, C.~B. Winkelmann, H.~Courtois, J.~P. Pekola, Thermal
  conductance of a single-electron transistor, Phys. Rev. Lett. 119~(7) (2017)
  077701.

\bibitem{sivre2018}
E.~Sivre, A.~Anthore, F.~D. Parmentier, A.~Cavanna, U.~Gennser, A.~Ouerghi,
  Y.~Jin, F.~Pierre, Heat {C}oulomb blockade of one ballistic channel, Nat.
  Phys. 14~(2) (2018) 145--148.

\bibitem{jezouin2013}
S.~J\'ezouin, F.~D. Parmentier, A.~Anthore, U.~Gennser, A.~Cavanna, Y.~Jin,
  F.~Pierre, Quantum limit of heat flow across a single electronic channel,
  Science 342~(6158) (2013) 601--604.

\bibitem{banerjee2018}
M.~Banerjee, M.~Heiblum, V.~Umansky, D.~E. Feldman, Y.~Oreg, A.~Stern,
  Observation of half-integer thermal {H}all conductance, Nature 559~(7713)
  (2018) 205--210.

\bibitem{zianni2007}
X.~Zianni, Coulomb oscillations in the electron thermal conductance of a dot in
  the linear regime, Phys. Rev. B 75~(4) (2007) 045344.

\bibitem{krawiec2006thermoelectric}
M.~Krawiec, K.~I. Wysoki{\'n}ski, Thermoelectric effects in strongly
  interacting quantum dot coupled to ferromagnetic leads, Phys. Rev. B 73~(7)
  (2006) 075307.

\bibitem{krawiec2007thermoelectric}
M.~Krawiec, K.~I. Wysokinski, Thermoelectric phenomena in a quantum dot
  asymmetrically coupled to external leads, Phys. Rev. B 75~(15) (2007) 155330.

\bibitem{murphy2008}
P.~Murphy, S.~Mukerjee, J.~Moore, Optimal thermoelectric figure of merit of a
  molecular junction, Phys. Rev. B 78~(16) (2008) 161406.

\bibitem{tsaousidou2010thermoelectric}
M.~Tsaousidou, G.~P. Triberis, Thermoelectric properties of a weakly coupled
  quantum dot: enhanced thermoelectric efficiency, J. Phys.: Cond. Matt.
  22~(35) (2010) 355304.

\bibitem{erdman2017thermoelectric}
P.~A. Erdman, F.~Mazza, R.~Bosisio, G.~Benenti, R.~Fazio, F.~Taddei,
  Thermoelectric properties of an interacting quantum dot based heat engine,
  Phys. Rev. B 95~(24) (2017) 245432.

\bibitem{dutta2020}
B.~Dutta, D.~Majidi, N.~W. Talarico, N.~Lo~Gullo, H.~Courtois, C.~B.
  Winkelmann, Single-quantum-dot heat valve, Phys. Rev. Lett. 125~(23) (2020)
  237701.

\bibitem{bjork2004}
M.~T. Bj{\"o}rk, C.~Thelander, A.~E. Hansen, L.~E. Jensen, M.~W. Larsson, L.~R.
  Wallenberg, L.~Samuelson, Few-electron quantum dots in nanowires, Nano Lett.
  4~(9) (2004) 1621--1625.

\bibitem{fasth2007direct}
C.~Fasth, A.~Fuhrer, L.~Samuelson, V.~N. Golovach, D.~Loss, Direct measurement
  of the spin-orbit interaction in a two-electron inas nanowire quantum dot,
  Physical review letters 98~(26) (2007) 266801.

\bibitem{wu2013}
P.~M. Wu, J.~Gooth, X.~Zianni, S.~F. Svensson, J.~G. Gluschke, K.~A. Dick,
  C.~Thelander, K.~Nielsch, H.~Linke, Large thermoelectric power factor
  enhancement observed in {I}n{A}s nanowires, Nano Lett. 13~(9) (2013)
  4080--4086.

\bibitem{roddaro2013}
S.~Roddaro, D.~Ercolani, M.~A. Safeen, S.~Suomalainen, F.~Rossella,
  F.~Giazotto, L.~Sorba, F.~Beltram, Giant thermovoltage in single {I}n{A}s
  nanowire field-effect transistors, Nano Lett. 13~(8) (2013) 3638--3642.

\bibitem{chen2018}
I.-J. Chen, A.~Burke, A.~Svilans, H.~Linke, C.~Thelander, Thermoelectric power
  factor limit of a 1{D} nanowire, Phys. Rev. Lett. 120~(17) (2018) 177703.

\bibitem{svilans2018}
A.~Svilans, M.~Josefsson, A.~M. Burke, S.~Fahlvik, C.~Thelander, H.~Linke,
  M.~Leijnse, Thermoelectric characterization of the {K}ondo resonance in
  nanowire quantum dots, Phys. Rev. Lett. 121~(20) (2018) 206801.

\bibitem{prete2019}
D.~Prete, P.~A. Erdman, V.~Demontis, V.~Zannier, D.~Ercolani, L.~Sorba,
  F.~Beltram, F.~Rossella, F.~Taddei, S.~Roddaro, Thermoelectric conversion at
  30 {K} in {I}n{A}s/{I}n{P} nanowire quantum dots, Nano Lett. 19~(5) (2019)
  3033--3039.

\bibitem{josefsson2018}
M.~Josefsson, A.~Svilans, A.~M. Burke, E.~A. Hoffmann, S.~Fahlvik,
  C.~Thelander, M.~Leijnse, H.~Linke, A quantum-dot heat engine operating close
  to the thermodynamic efficiency limits, Nat. Nanotechnol. 13~(10) (2018)
  920--924.

\bibitem{matthews2012}
J.~Matthews, E.~A. Hoffmann, C.~Weber, A.~Wacker, H.~Linke, Heat flow in
  {I}n{A}s/{I}n{P} heterostructure nanowires, Phys. Rev. B 86~(17) (2012)
  174302.

\bibitem{dick2010crystal}
K.~A. Dick, C.~Thelander, L.~Samuelson, P.~Caroff, Crystal phase engineering in
  single {I}n{A}s nanowires, Nano Lett. 10~(9) (2010) 3494--3499.

\bibitem{giazotto2006opportunities}
F.~Giazotto, T.~T. Heikkil{\"a}, A.~Luukanen, A.~M. Savin, J.~P. Pekola,
  Opportunities for mesoscopics in thermometry and refrigeration: Physics and
  applications, Rev. Mod. Phys. 78~(1) (2006) 217.

\bibitem{sivan1986multichannel}
U.~Sivan, Y.~Imry, Multichannel {L}andauer formula for thermoelectric transport
  with application to thermopower near the mobility edge, Phys. Rev. B 33~(1)
  (1986) 551.

\bibitem{davies1998physics}
J.~H. Davies, The physics of low-dimensional semiconductors: an introduction,
  Cambridge University Press, Cambridge, 1998.

\bibitem{mahan1996}
G.~D. Mahan, J.~O. Sofo, The best thermoelectric, Proc. Nat. Acad. Sci. 93~(15)
  (1996) 7436--7439.

\bibitem{lopez2013}
R.~L{\'o}pez, D.~S{\'a}nchez, Nonlinear heat transport in mesoscopic
  conductors: Rectification, {P}eltier effect, and {W}iedemann-{F}ranz law,
  Phys. Rev. B 88~(4) (2013) 045129.

\bibitem{prasad2004}
C.~Prasad, D.~K. Ferry, H.~H. Wieder, Energy relaxation studies in
  {I}n$_{0.52}${A}l$_{0.48}${A}s/
  {I}n$_{0.53}${G}a$_{0.47}${A}s/{I}n$_{0.52}${A}l$_{0.48}${A}s two-dimensional
  electron gases and quantum wires, Semicond. Sci. Technol. 19~(4) (2004) S60.

\bibitem{sugaya2002}
T.~Sugaya, J.~P. Bird, D.~K. Ferry, A.~Sergeev, V.~Mitin, K.-Y. Jang, M.~Ogura,
  Y.~Sugiyama, Experimental studies of the electron--phonon interaction in
  {I}n{G}a{A}s quantum wires, Appl. Phys. Lett. 81~(4) (2002) 727--729.

\end{thebibliography}

\clearpage

\widetext

\begin{center}
\textbf{\large Supplemental Material: Quantum confinement suppressing electronic heat flow below the Wiedemann-Franz law}
\end{center}
\makeatletter
\renewcommand{\theequation}{S\arabic{equation}}
\renewcommand{\thefigure}{S\arabic{figure}}
\global\long\def\theequation{S\arabic{equation}}
\global\long\def\thefigure{S\arabic{figure}}
\renewcommand{\thetable}{S\arabic{table}}

\bigskip
We provide here supplemental material and information concerning the sample fabrication process, details of the  electronic thermometry and heating, a description of the Landauer-B\"uttiker transport calculations, extended charge transport data of the quantum dot junction, and details of the thermal balance analysis.
\bigskip

\subsection{Sample Fabrication}

\begin{figure}[h]
      \vspace{.05cm}
	\includegraphics[width=0.5\columnwidth]{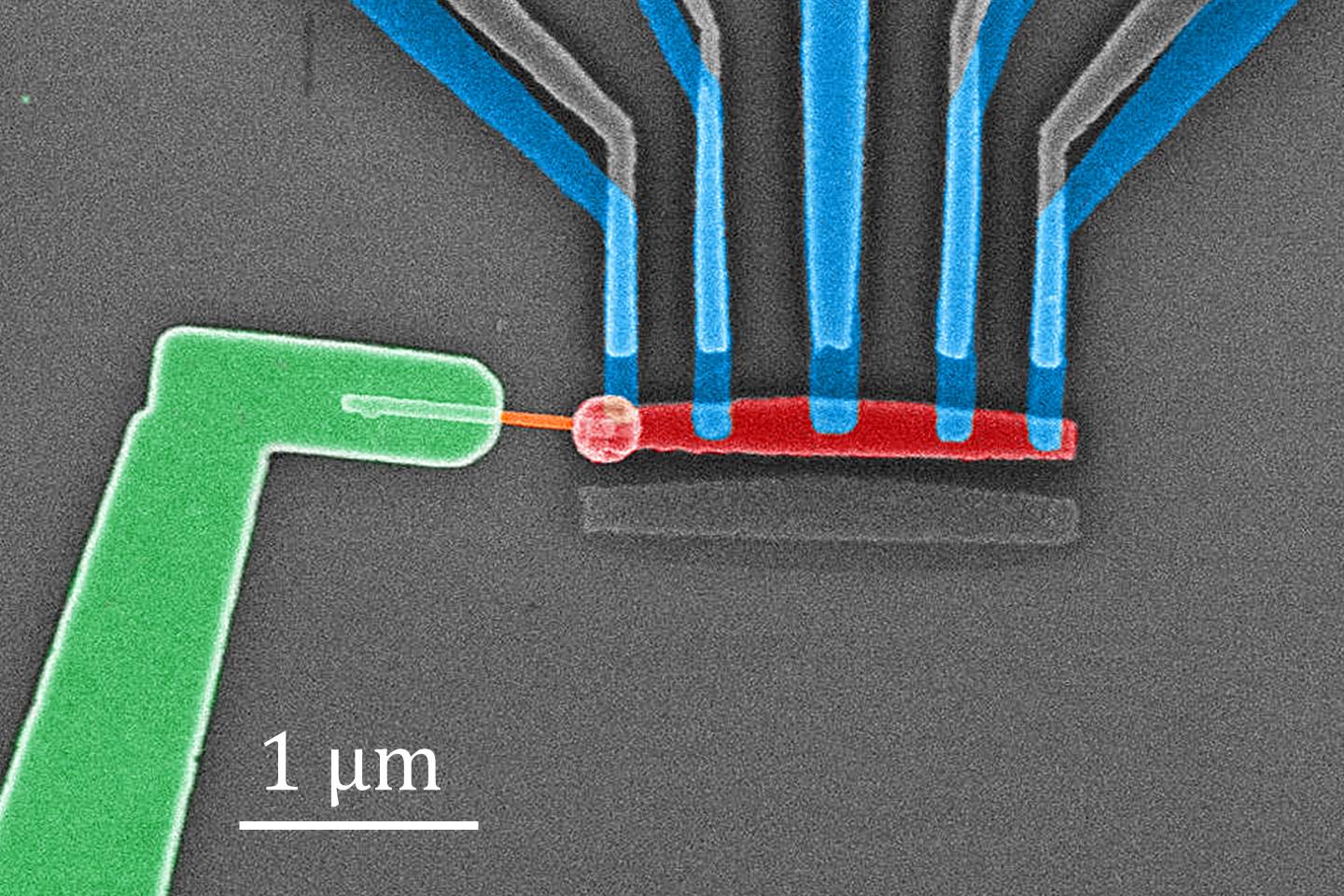}
	\caption{False-color scanning electron micrograph of the InAs nanowire device realized with two steps fabrication and shadow-evaporated Al-proximity junctions.}
	\label{fig:SI01}
\end{figure}

Our device consists of a InAs nanowire with a 70 nm diameter that was grown by chemical beam epitaxy seeded by a gold catalyst particle \cite{bjork2004, fasth2007direct}. The device is fabricated with two rounds of electron beam lithography (EBL) and subsequent hydrofluoric acid (HF) passivation and metal depositions. The substrate is a p-doped (resistivity 1 - 30 $\Omega $cm), single-side polished $2"$ Si wafer with 200 nm oxide allowing the underlying  Si substrate to be used as a global back gate. In the first lithography round, a bulky drain (shown in green in Fig. \ref{fig:SI01}) and part of the source (visible as a circle on the right hand side of the nanowire) are patterned. Subsequently, HF passivation (5 s in BOE 1:10 followed with fast rinsing in deionized water and immediate loading to the metal evaporator) is performed before contacts are metallized thermally as a stack of Ni (30 nm)/ Au (60 nm). The Ni layer helps with adhesion to the SiO$_2$ substrate and formation of an electrical contact to the nanowire.

Following a standard lift-off, a suspended P(MMA-MAA) copolymer based mask was spin-coated for the next EBL step where all the NIS junctions are defined as shown in Fig. \ref{fig:SI01} in blue and red color. After e-beam lithography, the mask is again loaded into an evaporator equipped with a tiltable sample holder. This allows fabricating both the normal-metal island (red) and the superconducting leads (blue) using the same mask in a single vacuum cycle. First, a 35 nm thick film of Al is deposited at $+16^\circ$ with respect to the evaporation source. It is indicated in Fig. \ref{fig:SI01} in blue color. To form the AlOx tunnel barriers for NIS probe tunnel junctions, the deposited Al layer is subjected to in-situ static oxidation immediately after the deposition is completed. This was accomplished by venting the chamber at air followed by an immediate pumping of the system. To complete the fabrication, a 60 nm thick Cu film was evaporated with the sample now tilted to $-16^\circ$ in the opposite direction compared to the Al deposition. This downwards-shifted copy of the mask pattern forms the source island (colored in red in Fig. \ref{fig:SI01}). The purpose of this Cu layer is to form the main part of the source electrode, connecting to the small source lead which was deposited in the first step. As a result of the two-angle evaporation through the same mask, two projections of the complete mask pattern will be formed on the substrate. The irrelevant, partially overlapping shadow copies of the various structures, evident in Fig. \ref{fig:SI01} are shown uncolored.

The leftmost Al electrode in Fig. \ref{fig:SI01} overlaps with the circular part of the source electrode made in the first lithography round. It therefore connects with a transparent contact to the source island without the oxidation, whereas the others connect via the oxide tunnel barrier through the Cu part of the source and hence display a large tunnel resistance. The leftmost lead allows for probing the charge transport of the nanowire. We do not observe any sign of a superconducting proximity effect on the nanowire caused by this electrode. This is most likely due to the fact that, in the absence of special care in cleaning the Au-Al interface {\it in vacuo}, residual contaminants such as a monolayer of water on the gold surface strongly reduce the interface transparency and thereby inhibit Andreev reflection. Further, the aluminum contact is much thinner (35 nm) than the nickel/gold island (90 nm) connecting the nanowire, which will further inhibit inducing superconducting correlations.

\subsection{$\mathrm{NIS}$ THERMOMETER AND HEATER CHARACTERIZATION }

 \begin{figure}[t]
      \vspace{.05cm}
	\includegraphics[width=0.7\columnwidth]{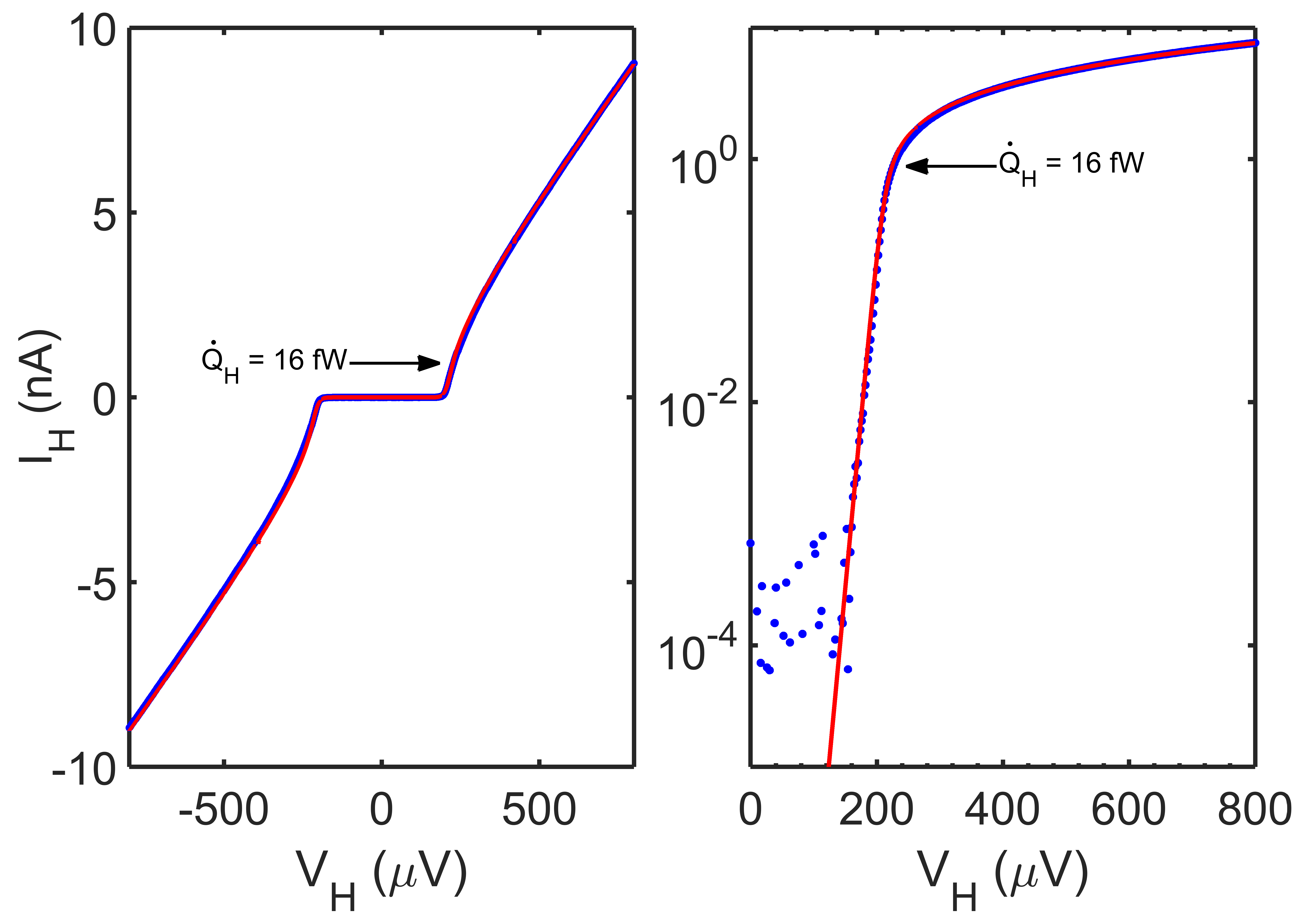}
	\caption{Current–voltage curve of the heater NIS junction using linear (left) and logarithmic (right) scale. Fit to Eq. (\ref{currentNIS}) is shown as red lines: $\Delta$ = 209 $\mu eV$, $R_T =$ 85.6 k$\Omega$ and $T_b$= 100 mK. A typical heater current level of $I_H$= 0.9 nA resulting in $\dot Q_H = 16$ fW and $\Delta T \approx 50$ mK (see Figs. 1 and 2 of the main article) is indicated with a black arrow.}
	\label{fig:SI_NIS_current}
\end{figure}

In this section, we describe the functioning of the NIS junctions as heaters and thermometers, and how their parameter values were extracted. The latter include the normal state tunnel resistance $R_T$, and the low-temperature superconducting energy gap $\Delta$. The I-V characteristics and heat current  through a single NIS junction reads as  \cite{giazotto2006opportunities}, respectively  
\begin{equation}\label{currentNIS}
\begin{array}{l}
  I= \frac{1}{2e R_{T}}\int_{-\infty}^{+\infty}  n_s(E,\Delta)\times [f_N(E-eV)-f_N(E+eV)] \ dE 

     \end{array}
\end{equation}
and 
\begin{equation}\label{heatNIS1}
\begin{array}{l}
  \dot{Q}_H (E) = \frac{1}{e^2 R_{T}}\int_{-\infty}^{+\infty} (E-eV) \ n_s(E,\Delta)\times [f_N(E-eV)-f_S(E)]  \ dE 
                               
     \end{array}
\end{equation}
where $n_S(E)$ is the normalized BCS density of states with an energy gap of $\Delta$, and $f_S(E)$ and $f_N(E)$ are the quasi-particle occupation factors for the superconductor and normal metal. It is worth mentioning that the charge current depends on the electronic temperature of the normal metal but not on the temperature of the superconductor for temperatures well below $\Delta/k_B$. Despite the apparent simplicity of the above expressions, they produce quantitatively correct predictions in most experimentally interesting cases.
 The low-temperature experimental characteristic of the heater NIS junction of our device at $T_b=100$ mK is shown in Fig. \ref{fig:SI_NIS_current}, both on linear and logarithmic scale, together with the theoretical $I - V$ characteristic using Eq. (\ref{currentNIS}). We see that the theory line catches all features except at the lowest currents where the noise of the current preamplifier contributes to the scatter of the data points.

 \begin{figure}[t!]
     \vspace{.05cm}
	\includegraphics[width=0.6\columnwidth]{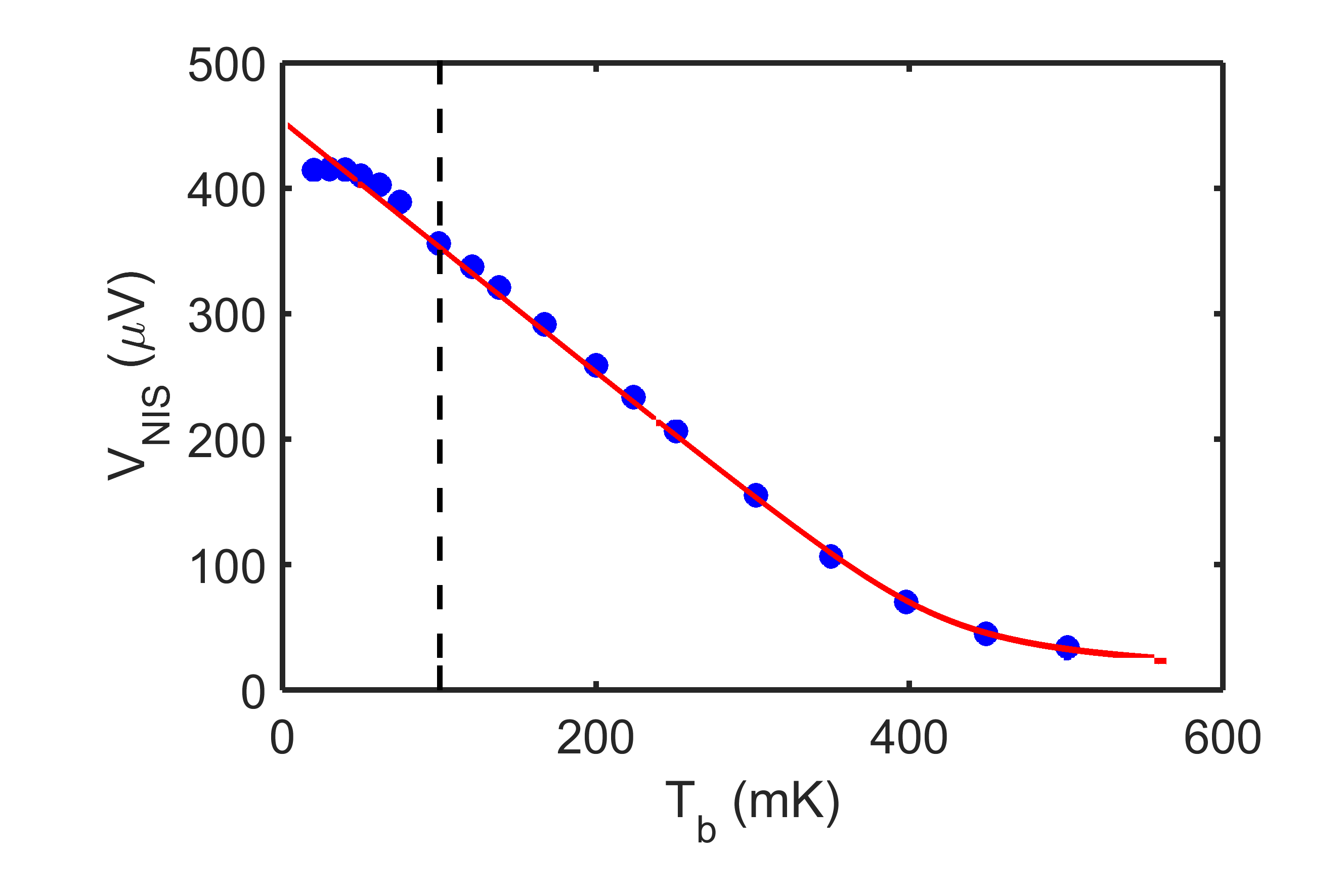}
	\caption{Measured voltage across the thermometer junctions with a floating current of $I_\mathrm{NIS}=$ 5 pA as a function of the bath temperature $T_b$. The dashed line presents the bath temperature $T_b = 100$ mK used in the heat flow measurements. The red line shows the expected response of Eq. (\ref{currentNIS}) with $\Delta = 209~\mu$eV and $R_{T1} = R_{T2} \sim 96.2~k\Omega$ obtained from a similar fit as in Fig. \ref{fig:SI_NIS_current}.}
	\label{fig:SI_Calibration}
\end{figure}

The heating of the source island was made by applying a voltage $V_H$ as shown in Fig. 1a of the main article. The ensuing power $\dot Q_H$ is given by Eq. (\ref{heatNIS1}). A peculiar property of the NIS junctions is that for bias voltages slightly less than the gap voltage $\Delta /e$, one can find a region where $\dot{Q}_H$ is negative, i. e., the normal electrode is cooled. This Peltier effect has been extensively studied \cite{giazotto2006opportunities}. Due to uncertainties in the precise determination of the power at sub-gap voltages, we however have not included data for negative $\dot{Q}_H $ in our heat balance analysis but rather focus on the heating side.

 Thermometry is performed by using two NIS junctions in series (SINIS). We bias the pair of the NIS junctions with a constant current of $I_\mathrm{NIS}=$ 5 pA and measure the voltage drop $V_\mathrm{NIS}$ across the junctions to determine the electron temperature of the source island~\cite{giazotto2006opportunities}. The thermometer is calibrated by varying the bath temperature $T_b$ of the cryostat. The calibration is done at equilibrium without heating the source island, so that the electronic temperature of the source follows the cryostat temperature (equal to the substrate phonon temperature) and results in the response presented in Fig. \ref{fig:SI_Calibration}. The voltage $V_\mathrm{NIS}$ changes as a result of thermal excitations on the normal metal lowering the voltage from the low temperature threshold value corresponding to approximately the superconductor gap $\Delta/e \approx 200~\mu$V per junction. At low $T_b < 50$ mK, we indeed see the saturation at $V_\mathrm{NIS} \approx 400~\mu$V. All our measurements are performed at $T_e \geq 100$ mK, making sure that the thermometer operates well above the low temperature saturation.

\subsection{Analysis of transport through a single quantum dot orbital}
 
\begin{figure}[t]
\centering
	\includegraphics[width=0.45\columnwidth]{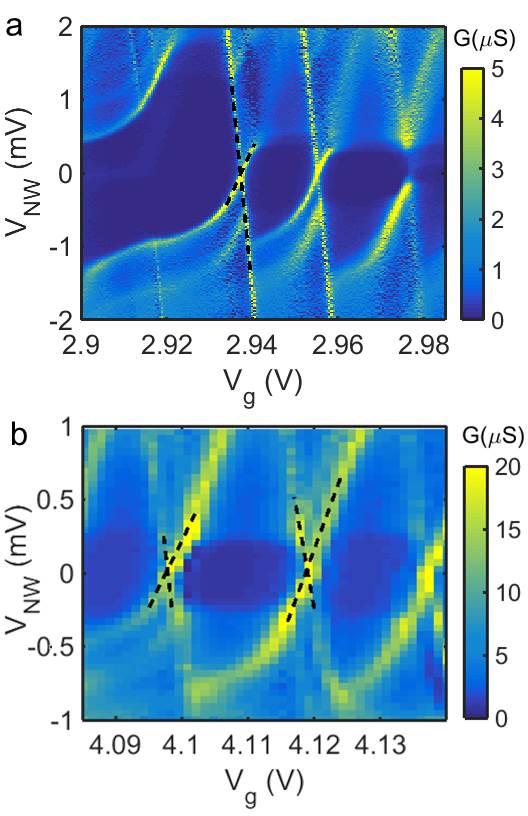}
		\caption{Differential conductance maps near $V_{g} \sim 2.9 $ V (raw data) (a) and $V_{g} \sim 4.1 $ V (interpolated data) (b). The bath temperature $T_b$ is 100 mK. 
		}
	\label{fig:Stability}
\end{figure}

As evidenced by the conductance peaks (see figures in the main paper) and by the stability diagrams (Fig.~\ref{fig:Stability}) a quantum dot forms in the nanowire below $V_g=4.5$ V. From the stability diagrams we find charging energies $E_c\approx 1.5-2$ meV. Additional sequential tunneling and cotunneling resonances show that, in contrast with metallic islands, there is also a substantial energy splitting between the quantized quantum dot orbitals, $\Delta \varepsilon \gg \gamma, kT$.

The energy scales of our device in the quantum dot regime, with $E_c\gg \gamma>k_BT$, unfortunately prevents the use of theoretical approaches based on perturbation theory in $\gamma$, and make theories that include $E_c$ on an approximate footing unreliable. Our approach is instead to model the QD with a non-interacting model close to the charge-degeneracy points (Coulomb peaks). In this case the currents through a \textit{single} quantum dot orbital can be calculated using Landauer-B\"{u}ttiker transport theory as \cite{sivan1986multichannel, davies1998physics}, 
\begin{equation}
    I=\frac{2e}{h}\int_{-\infty} ^\infty {\mathcal T}(E)\cdot \Delta f\ dE,
    \label{Eq:I}
\end{equation}
\begin{equation}
    \dot{Q_e}=\frac{2}{h}\int_{-\infty} ^\infty (E-\mu_s)\cdot {\mathcal T}(E)\cdot \Delta f\ dE,
    \label{Eq:Q}
\end{equation}
where
\begin{equation}
    \Delta f = f_s - f_d, \ \ f_n = \left ( \exp{\left(\frac{E-\mu_n}{k_bT_n}\right)} + 1 \right)^{-1},
    \label{Eq:fermi}
\end{equation}
and
\begin{equation}
   {\mathcal T}(E)=\frac{4\gamma_s\gamma_d}{\gamma^2}\frac{\left( \frac{\gamma}{2}\right )^2}{(E - (\varepsilon - e\alpha V_g))^2 + \left( \frac{\gamma}{2}\right )^2},
    \label{Eq:trans}
\end{equation}
with $\gamma=\gamma_s + \gamma_d$. In Eq.~(\ref{Eq:trans}) we assumed that $\gamma_{s,d}$ are energy-independent. Note that we above analysis allows finding the values of the pair $(\gamma_{s},\gamma_{d})$, but does not allow assigning which one is which. We therefore list the tunnel couplings as $\gamma_{1,2}$ from hereon, without specifying which one is $\gamma_{s}$ and $\gamma_{d}$, respectively.

The relevant physical quantities for this study are the linear electrical and thermal conductances, $G=\left. \frac{\partial I}{\partial V_{NW}} \right|_{\Delta T=0}$ and $\kappa_e=\left.\frac{\partial \dot{Q_e}}{\partial (\Delta T)}\right|_{I=0}$, which are obtained by differentiating Eqs.~(\ref{Eq:I},\ref{Eq:Q}).

\subsection{Quantum dot characterisation}

In the quantum dot regime (at small $V_{g}$), the conductance map provides all information needed to  determine the parameters of the quantum dot. Fig. \ref{fig:Stability} displays the measured differential conductance maps (obtained by numerical differentiation) as a function of both the bias and gate voltages $V_{NW}$ and $V_g$, respectively, around the operation points of Figs. 2 and 3 of the main article. Coulomb diamonds can be seen clearly and the charging energy of $E_c \sim$ $1.5-2$ meV is estimated from extrapolating the bias level to the top of a diamond.

The positive slope of the Coulomb diamonds in Fig. \ref{fig:Stability} is given by $\beta = \frac{C_g}{C_d+C_g}$ and the negative one is given by $\beta' = \frac{C_g}{C_s}$. Here the $C_s,C_d,C_g$ are the capacitances between the dot and source, drain and the gate respectively. The total capacitance of the dot to the outside world is the sum of all capacitances as $C_{\Sigma} = C_s+C_d+C_g$. An important parameter in the calculations is the lever arm defined by the ratio $\alpha = \frac  {C_g}{C_{\Sigma}}$ as it translates changes in gate voltage to energy changes for electrons on the quantum dot, $\Delta \varepsilon = -e\alpha\Delta V_g$. From the conductance maps close to $V_g \sim$ 4.1 V and $V_g \sim$ 3 V  we extract a local lever arm $\alpha_m$ from the slopes of conductance lines. We observe that the level arm value varies slightly with the gate voltage, see Tab.\ref{Tab:lever_arm}. In order to account for uncertainties in the determination of gate coupling, we also performed the full theoretical analysis at $\alpha_m\pm0.01$ for $V_g \sim 3$ V  and $\alpha_m\pm0.02$ for $V_g \sim  4.1$ V.

\begin{figure}[t]
     \vspace{.05cm}
	\includegraphics[width=0.6\columnwidth]{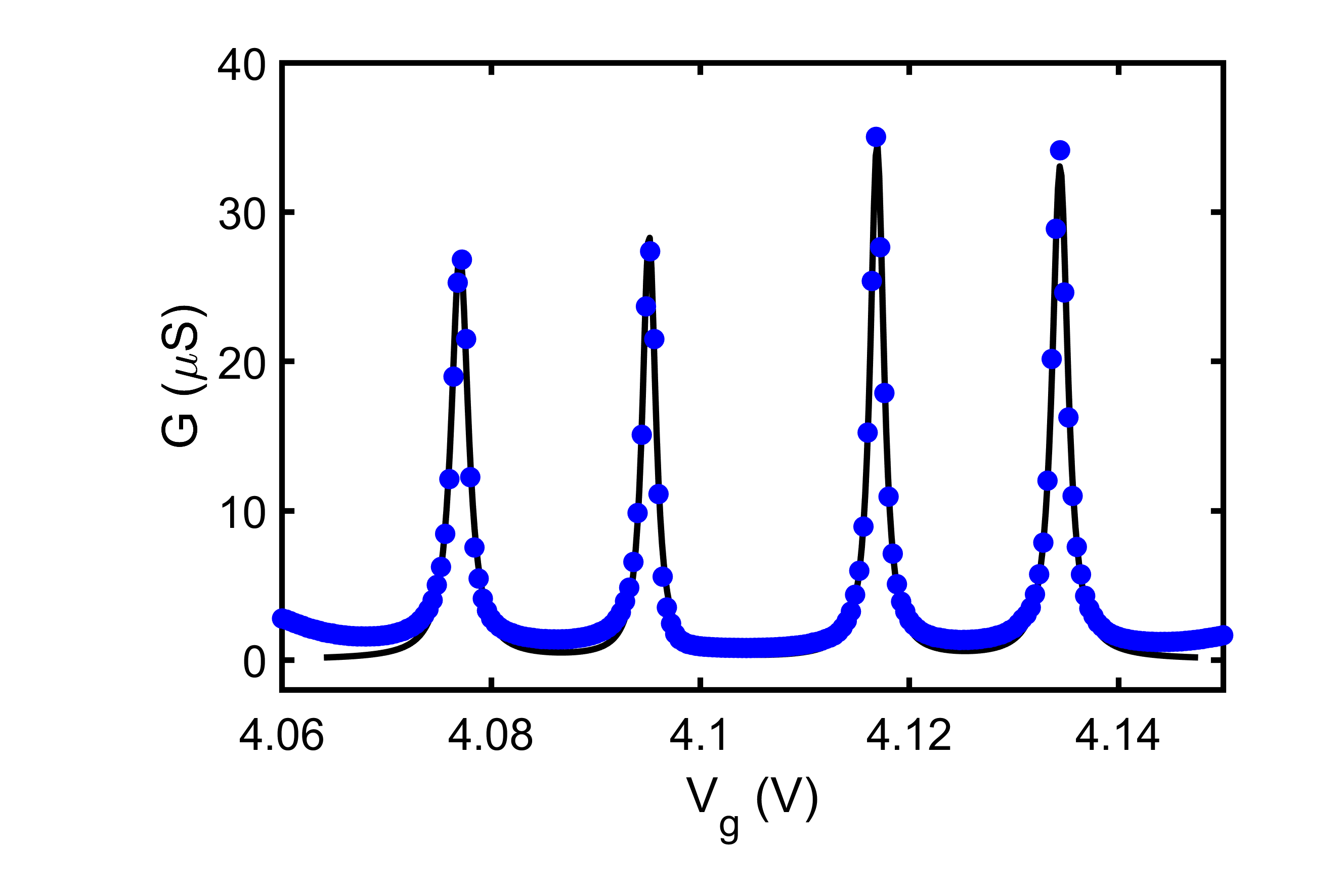}
	\caption{Measured (blue markers) and calculated (solid line) charge conductance of the device around the charge degeneracy points close to 4.1 V. The full transmission function used for the theory prediction is obtained by combining one $\mathcal{T}(E)$ for each peak, determined by fitting the calculated $G$ to the measured data in the vicinity of a single peak.}
	\label{fig:SI_G_4V}
\end{figure}

\begin{table}
\centering
    \vspace{.4cm}
    \begin{tabular}{||c|c|c||}
        \hline
        $V_g$ = $2.933$ V  &  $V_g$ = $4.095$ V   &  $V_g$ = $4.117$ V \\ 
         \hline
         $\alpha_l$ = 0.095  &$\alpha_l$ =  0.06& $\alpha_l$ =  0.06 \\
         $\alpha_m$ = 0.105 &$\alpha_m$ = 0.08&$\alpha_m$ = 0.08\\
          $\alpha_u$ = 0.115 &$\alpha_u$ = 0.1&$\alpha_u$ = 0.1\\
         \hline

    \end{tabular}
    \caption{\label{Tab:lever_arm} Extracted lower ($\alpha_l$) and  upper ($\alpha_u$) bounds of gate couplings for the Coulomb peaks, and their mean value ($\alpha_m$) at the resonances considered in the main article.}
\end{table}
Next we extract a pair of tunnel couplings $(\gamma_1, \gamma_2)$ for each Coulomb peak by fitting the calculated zero-bias conductance (obtained using Eqs.~\ref{Eq:I} and~\ref{Eq:trans}) as a function of the gate voltage to the measured counterpart. The parameter values for the two tunnel couplings are obtained uniquely from the height $\frac{\gamma_1\gamma_2}{\gamma_1+\gamma_2}$ and the width $ \gamma_1+\gamma_2$ of the transmission function, which corresponds roughly to the height and width of the Coulomb peak. Since we have already determined the appropriate $\alpha$ this fitting process involves no additional fitting parameters.  When performing these fits, we restrict the $G$ data to only cover a single Coulomb peak and use $T_b = T_s = T_d =100$ mK, which was the temperature of the device during the measurement of $G$. The resulting best fits obtained using $\alpha_m$ are shown in Fig. 2 of the main paper for $V_g\sim 3$ V and in Fig.~\ref{fig:SI_G_4V} for the resonances at $V_g\sim$ 4.1 V corresponding to the data presented in Fig. 3 of the main paper. The extracted values are shown in Tab.~\ref{Tab:Gammas}.  

Around 4.1 V (Fig. 3 of the main article), the total transmission function ${\mathcal T}(E)$ of the device is taken to be the sum of the four individual Lorentzian transmission functions for each Coulomb peak, centered around the $V_g$ values listed in Tab.~\ref{Tab:Gammas}.

\begin{table}
\centering
    \vspace{.4cm}
 \begin{tabular}{||c |c |c | c||}
        \hline
        $V_g$ (V) &$\alpha$    &$\gamma_1$ ($\mu$eV) & $\gamma_2$ ($\mu$eV) \\ 
         \hline
        2.938 & $\alpha_m$ = 0.105 & 55.5& 2.1  \\
         \hline
         4.077  &$\alpha_m$ = 0.08 & 137.5 &15.5 \\
        \hline
         4.095  &$\alpha_m$ = 0.08 & 105.9 &12.4  \\
        \hline
         4.117  &$\alpha_m$ = 0.08 & 104.8&15.9  \\
        \hline
         4.13       &$\alpha_m$ = 0.08 & 122.9&17.5  \\
        \hline

    \end{tabular}
    \caption{\label{Tab:Gammas}Extracted tunnel couplings for the Coulomb peaks in two regimes $V_{g} \sim 2.9$ V, and $V_{g} \sim 4.1$ V.}
\end{table}

\begin{figure}
	\includegraphics[width=0.6\columnwidth]{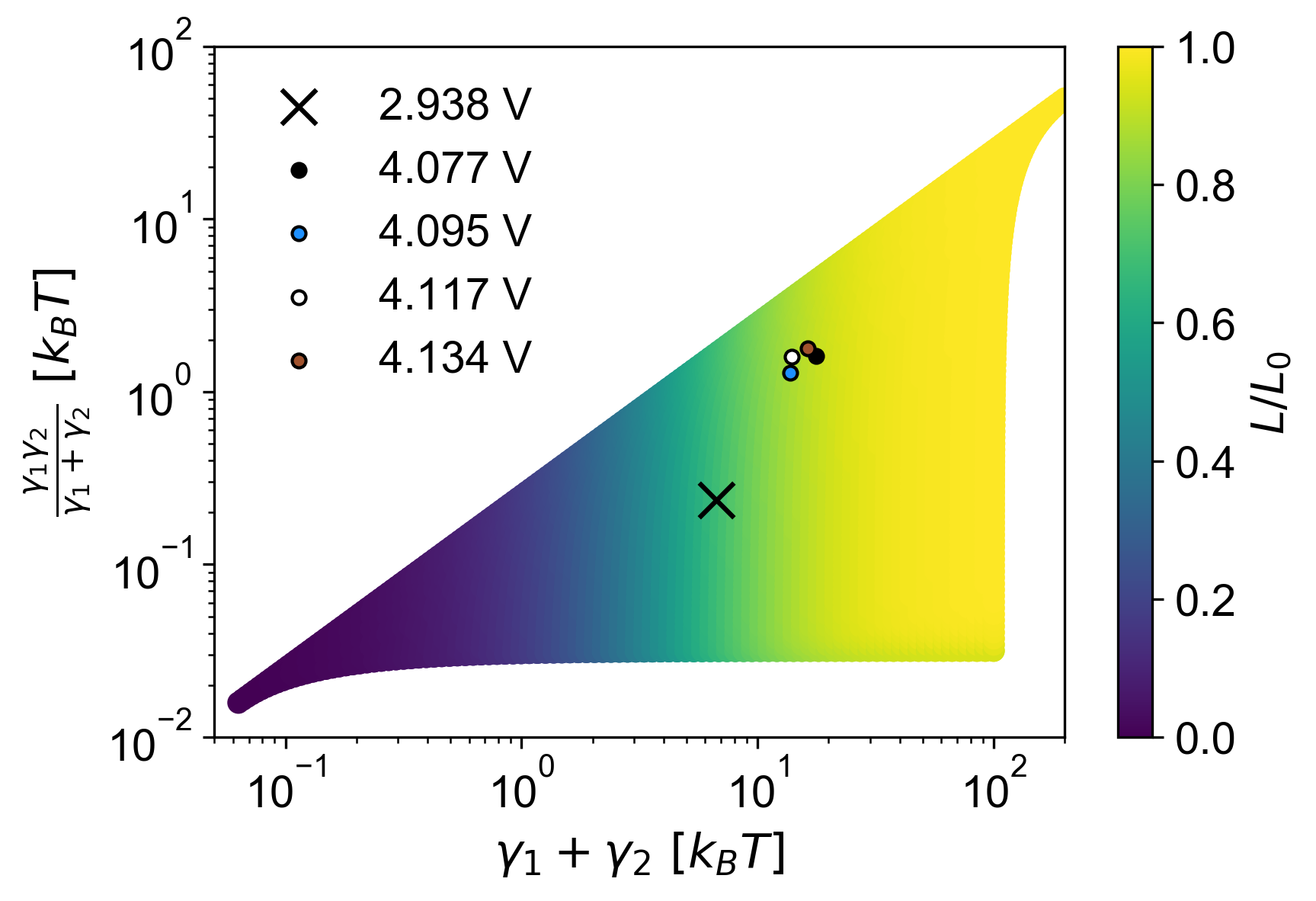}
	\caption{Calculated $L/L_0$ on resonance as a function of the width ($\gamma_1+\gamma_2$) and amplitude ($\frac{\gamma_1\gamma_2}{\gamma_1+\gamma_2}$) of $\mathcal{T}$(E). Markers show the theoretical predictions for the resonances studied in the main manuscript, as indicated by the legend.}
	\label{fig:scaling}
\end{figure}

\begin{figure}[h!]
     \vspace{.05cm}
	\includegraphics[width=0.6\columnwidth]{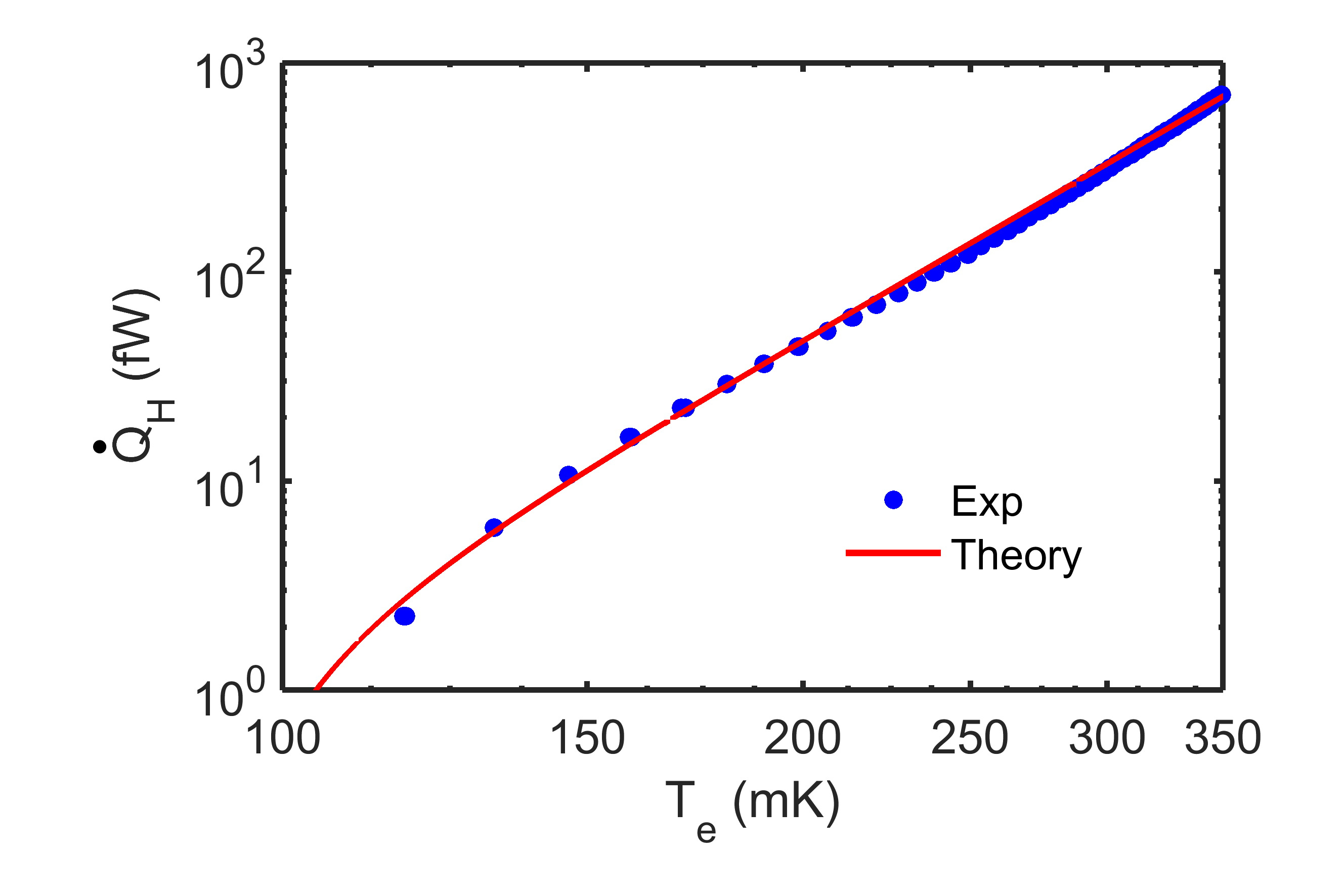}
	\caption{Heating power $\dot Q_H$ applied to the source island as a function of the measured source electron temperature $T_e$ at $V_g = 0$ V and  $T_b$ = 100 mK. The red curve is a fit, see text.}
	\label{fig:SI_eph_Cu}
\end{figure}

\begin{figure*}[t!]
    \vspace{.05cm}
	\includegraphics[width=0.75\columnwidth]{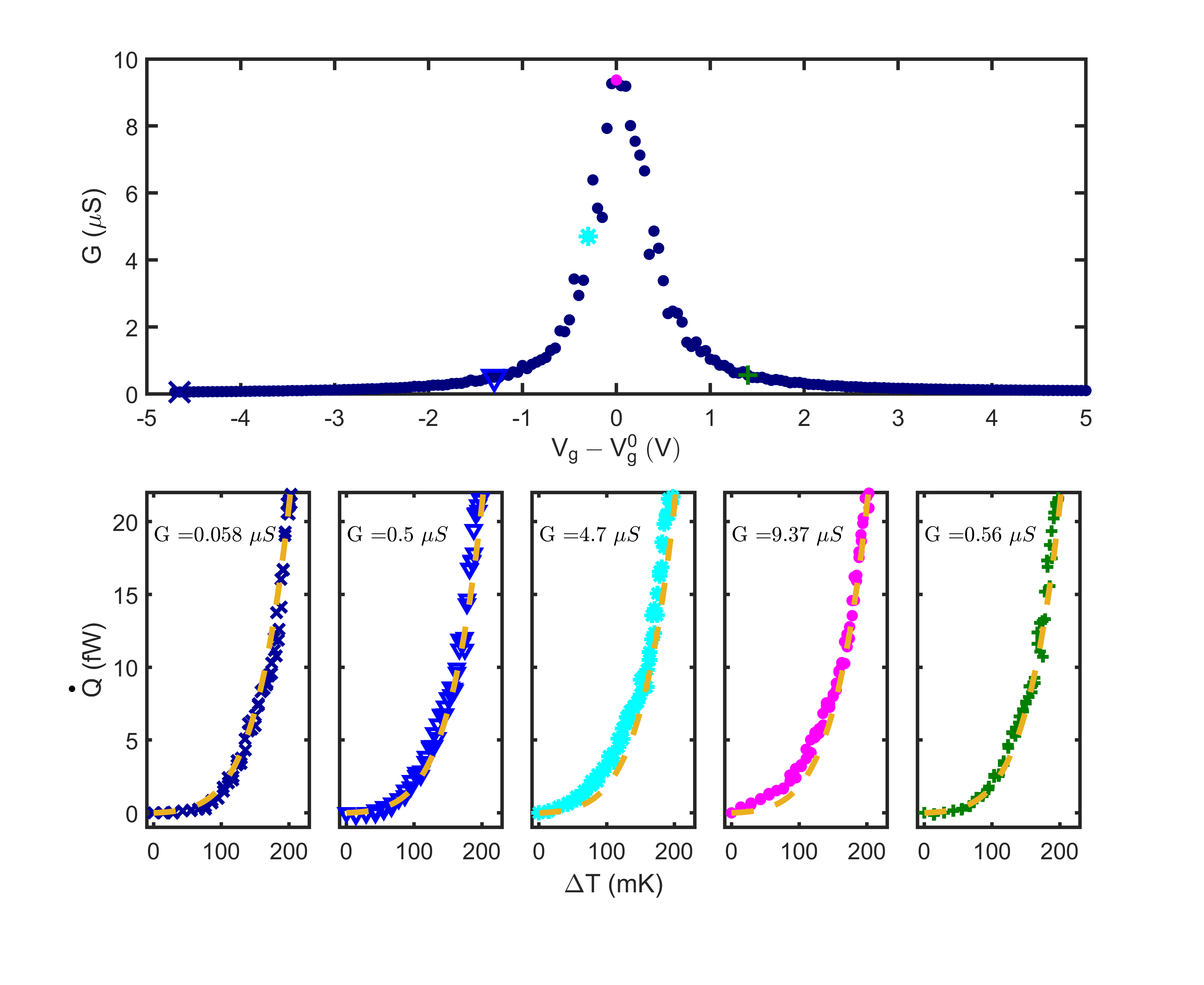}
	    \vspace{-1.1cm}
	\caption{Top : Charge conductance $G$ peak around the resonance at ${V_g}^0= 2.938$ V. Bottom : heat flow $\dot{Q}$ as a function of the temperature difference at several values of the gate potential indicated by color symbols in the bottom panel. The dashed line is the best $\propto (T_e^6-T_b^6)$ fit obtained from the data in the leftmost sub-panel, and displayed identically in all sub-panels for reference.}
	\label{fig:SI_e-ph}
\end{figure*}

In addition to modelling the device at the  operating conditions of the experiment, the theory also allows us to investigate how the Wiedemann-Franz law violations scale with system parameters. Focusing on the resonant condition, i.e. gating the device to the middle of a conductance peak, we calculate how $L/L_0$ scales with the width ($\gamma_1+\gamma_2)$ and amplitude ($\frac{\gamma_1\gamma_2}{\gamma_1+\gamma_2}$) of $\mathcal{T}(E)$. The result is shown in Fig.~\ref{fig:scaling}, where the theory predictions for the five resonances analyzed in this work are also highlighted. From the figure it is evident that there are two paths for decreasing the $L/L_0$ ratio: lowering $\gamma_1+\gamma_2$ or increasing $k_BT$. However, lowering the tunnel couplings is non-trivial in our device as the quantum dot forms spontaneously at low carrier concentrations and there is little experimental control over the coupling strength. In addition, an effect of overall lower values of $\gamma_1$ and $\gamma_2$ is a reduced heat flow, which can be hard to detect experimentally since the signal is more easily swallowed by the noise-floor. The other approach, to increase $k_BT$, also has its limitations as the NIS thermometer requires the Al leads to be well below the critical temperature of the superconductor and the e-ph coupling of the source island to be small \cite{giazotto2006opportunities}. One can thus conclude that the present device is very well suited for studying violations of the Wiedemann-Franz law due to quantum confinement given the constraints of the state-of-art technologies used in the study.

\subsection{Determination of the heat flows, analysis of the electron-phonon couplings}

The relation ${\dot Q}_H(T_e, V_g=0)$ between the applied heating power and the source island electronic temperature at $V_g=0$ is shown in Fig. \ref{fig:SI_eph_Cu}.  The good quantitative agreement with an electron-phonon type thermal law \cite{giazotto2006opportunities} shows that electron-phonon coupling must be the dominant thermal leakage channel out of the source island, in the absence of electronic heat conduction through the nanowire. The red curve is a fit with $\dot Q_H = \Sigma\Omega ({T_e}^5- {T_b}^5)$. By using the geometrically estimated total volume $\Omega=4.26\pm0.2 \times10^{-20} $ m$^3$ of the source island, we obtain the fitted value $\Sigma=2.5 \pm 0.1\times 10^9 $ Wm$^{-3}$K$^{-5}$ of the average electron-phonon coupling coefficient in the metallic source island, in good agreement with the expected coupling coefficients of Cu and Au \cite{giazotto2006opportunities}.

At a given resonance peak, the electronic heat conductance is experimentally determined by subtracting a local reference of heat $\dot Q$ measured close to the resonance at a point where the electronic contribution is negligible. Fig. \ref{fig:SI_e-ph} shows that the heat flow is constant within noise at low electrical conductance $G$ far enough away from the main peak although $G$ changes by one order of magnitude. Only close to the resonance peak, we observe an extra contribution identified as the electronic part. On both sides of the peak, $\dot Q$ has the same background level when $G$ is small enough. It is also worth noting that the background variation between $V_g = 0$ and $V_g = 3$ V is vanishingly small for $\Delta T < 40$ mK as seen from Fig. 2c of the main article. For small $\Delta T$, the electronic contribution dominates.

\begin{figure}[t!]
	\includegraphics[width=0.6\columnwidth]{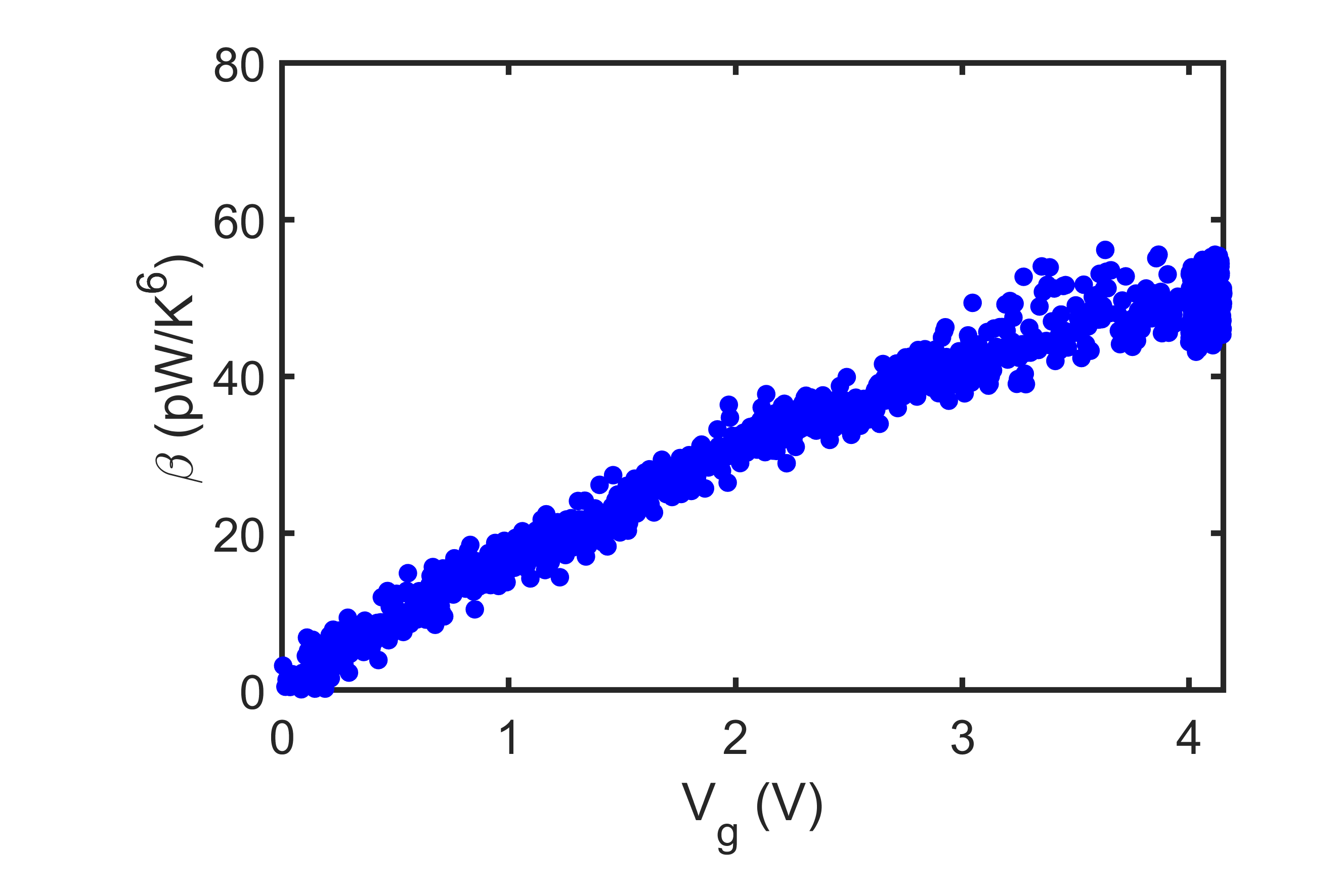}
	\caption{Gate dependence of the electron-phonon coupling: $\beta$ is extracted by fitting ${\dot Q}(T_e,V_g)$  at each $V_g$ (excluding conduction resonances) with a $\beta(T_e^6-T_b^6)$ power law.}
	\label{fig:SI_eph_NW}
\end{figure}

In order to understand this background contribution, we have analysed the ${\dot Q}(T_e,V_g)$ curves in the entire non-conducting regime of the nanowire. For this purpose, we have focused on the regime between $V_g=0\hdots 4.5$ V, excluding conduction resonances, that is, data sets at values of $V_g$ at which $G>0.5\ \mu$S. 
The background part of the heat flow increases steeply at $\Delta T \gtrsim T_b$ and is related to electron-phonon coupling in the nanowire. A $(T_e^6-T_b^6)$ law provided by far the best agreement. Because on the microscopic level the e-ph coupling can be quite different in InAs and the metallic island, it is not surprising that we observe a different exponent for the e-ph coupling of both systems \cite{giazotto2006opportunities}.

The prefactor $\beta$ is plotted in Fig. \ref{fig:SI_eph_NW}. As expected, $\beta$ increases smoothly with $V_g$, supporting the hypothesis of a dependence on the carrier concentration in a segment of the nanowire not belonging to the quantum dot. This could be for example the portion of the nanowire underneath the source island, of volume $V=7.7\times10^{-22}$ m$^3$. Making this assumption, the e-ph heat conductance per unit volume in the metallic source is on the same order of magnitude as that of the nanowire. Note that our method is probably underestimating $\beta$ by a constant shift, since we assumed its value to be 0 at $V_g=0$ V.

\end{document}